%% file: paper.tex
\documentclass[sigconf, nonacm, screen, table, breaklinks,colorlinks=true,linkcolor=blue,anchorcolor=black,citecolor=black,filecolor=black,menucolor=black,runcolor=black,urlcolor=blue, 10pt]{acmart}
\usepackage[abbreviations]{foreign}
\AtBeginDocument{%
  \providecommand\BibTeX{{%
    \normalfont B\kern-0.5em{\scshape i\kern-0.25em b}\kern-0.8em\TeX}}}
\usepackage[utf8]{inputenc}
\usepackage[T1]{fontenc}
\usepackage{xcolor}
\usepackage{algorithmic}
\usepackage{graphicx}
\usepackage{textcomp}
\usepackage{listings}
\usepackage{url}
\usepackage{pifont}
\usepackage{booktabs}
\usepackage{multirow}
\usepackage[inline]{enumitem}
\usepackage{hyperref}
\usepackage{subcaption}
\usepackage[margin=0pt,skip=6pt,belowskip=0pt,font={small,stretch=0.9},labelfont=bf]{caption}
\usepackage{tabularx, makecell, booktabs}
\usepackage{tikz}
\definecolor{ashgrey}{rgb}{0.252,0.751,0.452}
\usepackage{framed}
\colorlet{shadecolor}{ashgrey}
\usepackage[font={small}]{caption}
\usepackage[T1]{fontenc}
\usepackage{underscore}
\usepackage{graphicx}

\newcommand{\cmark}{\ding{51}}
\newcommand{\xmark}{\ding{55}}
\usepackage{setspace}
\usepackage{centernot}
\usepackage{pifont}
\usepackage[subtle]{savetrees}

\usepackage{caption}
\usepackage{balance}
\usepackage[export]{adjustbox}
\usepackage{eso-pic}
\usepackage[most]{tcolorbox}
\usepackage{eso-pic} 
\begin{document}

\title[Practical Forecasting of Cryptocoins Timeseries using Correlation Patterns]{Practical Forecasting of Cryptocoins Timeseries\\using Correlation Patterns}

\author{Pasquale De Rosa}
\affiliation{
	\institution{University of Neuch\^{a}tel}
	\country{Switzerland}}
\email{pasquale.derosa@unine.ch}
\orcid{0000-0001-9726-7075}

\author{Pascal Felber}
\affiliation{
	\institution{University of Neuch\^{a}tel}
	\country{Switzerland}}
\email{pascal.felber@unine.ch}
\orcid{0000-0003-1574-6721}

\author{Valerio Schiavoni}
\affiliation{
	\institution{University of Neuch\^{a}tel}
	\country{Switzerland}}
\email{valerio.schiavoni@unine.ch}
\orcid{0000-0003-1493-6603}

\begin{abstract}
	Cryptocoins (\ie Bitcoin, Ether, Litecoin) are tradable digital assets.
	Ownerships of cryptocoins are registered on distributed ledgers (\ie, blockchains).
	Secure encryption techniques guarantee the security of the transactions (transfers of coins among owners), registered into the ledger.
	Cryptocoins are exchanged for specific trading prices.
	The extreme volatility of such trading prices across all different sets of crypto-assets remains undisputed.
	However, the relations between the trading prices across different cryptocoins remains largely unexplored.
	Major coin exchanges indicate trend correlation to advise for sells or buys. 
	However, price correlations remain largely unexplored.
	We shed some light on the trend correlations across a large variety of cryptocoins, by investigating their coin/price correlation trends over the past two years. 
	We study the causality between the trends, and exploit the derived correlations to understand the accuracy of state-of-the-art forecasting techniques for time series modeling (\eg, GBMs, LSTM and GRU) of correlated cryptocoins.
	Our evaluation shows \emph{(i)} strong correlation patterns between the most traded coins (\eg, Bitcoin and Ether) and other types of cryptocurrencies, and \emph{(ii)} state-of-the-art time series forecasting algorithms can be used to forecast cryptocoins price trends.
	We released datasets and code to reproduce our analysis to the research community.
\end{abstract}


\maketitle

\input{authors-version}
\input{intro}
\input{background}
\input{dataset}
\input{correl}
\input{causality}
\input{eval}

\input{discussion}
\input{relwork}
\input{conclusion}

\bibliographystyle{ACM-Reference-Format}
\bibliography{biblio}
\end{document}

%% file: authors-version.tex

\def\confname{17th ACM International Conference on Distributed and Event-based Systems (DEBS'23)}
\def\confyear{2023}
\def\confdoi{XXX}

\definecolor{yellowPaper}{HTML}{fff8ae}
\AddToShipoutPictureFG*{%
  \AtTextUpperLeft{%
    \adjustbox{raise=30pt}{
    \begin{tcolorbox}[width=1\textwidth,colback=yellowPaper,enhanced,frame hidden,sharp corners]  
        \centering\scriptsize
        \copyright~\confyear\ 	
         by the Association for Computing Machinery, Inc. (ACM). Permission to make digital or hard copies of portions of this work for personal or classroom use is granted without fee provided that the copies are not made or distributed for profit or commercial advantage and that copies bear this notice and the full citation on the first page in print or the first screen in digital media. Copyrights for components of this work owned by others than ACM must be honored. Abstracting with credit is permitted.
        This is the author's version of the work.
        The final authenticated version is available online at \href{https://doi.org/10.1145/3583678.3596888}{https://doi.org/10.1145/3583678.3596888} 
        and has been published in the proceedings of the 
        \confname.
     \end{tcolorbox}} 
  }%
}%

\hypersetup{
    pdfcopyright={\copyright~\confyear\  Copyright 2023 by the Association for Computing Machinery, Inc. (ACM). Permission to make digital or hard copies of portions of this work for personal or classroom use is granted without fee provided that the copies are not made or distributed for profit or commercial advantage and that copies bear this notice and the full citation on the first page in print or the first screen in digital media. Copyrights for components of this work owned by others than ACM must be honored. Abstracting with credit is permitted.
    	This is the author's version of the work.
    	The final authenticated version is available online at \href{https://doi.org/10.1145/3583678.3596888}{https://doi.org/10.1145/3583678.3596888} 
    	and has been published in the proceedings of the 
    	\confname.}
}

%% file: intro.tex
\section{Introduction}

Cryptocurrencies (\ie cryptocoins) are tradable digital assets backed by secure encryption techniques to ensure the security of transactions (typically, the transfer of coins across wallets).
Examples include Bitcoin~\cite{nakamoto2008bitcoin}, Ether (the native cryptocurrency of the Ethereum blockchain~\cite{buterin2013ethereum}) Litecoin (a fork of the original Bitcoin network), and many more (CoinMarketCap~\cite{coinmarketcap} lists 9144 coins in November 2022).
Cryptocoins are traded as digital money: the first useful Bitcoin transaction was used for a peer-to-peer payment by Satoshi Nakamoto in 2009 (\url{https://www.blockchain.com/btc/block/170}).
Cryptocoins are traded online on centralized or decentralized \emph{exchange} platforms (\eg Coinbase~\cite{coinbase}, Kraken~\cite{kraken}, Binance~\cite{binance}, Uniswap~\cite{uniswap}, \etc).
With a current estimated worldwide market capitalization of 1.71 trillion dollars, the cryptocoin economy roughly match the GDP of South Korean in 2021~\cite{IMF}.

\begin{figure}[!t]
	\begin{center}
		\includegraphics[scale=0.68,trim={0 0 0 0}]{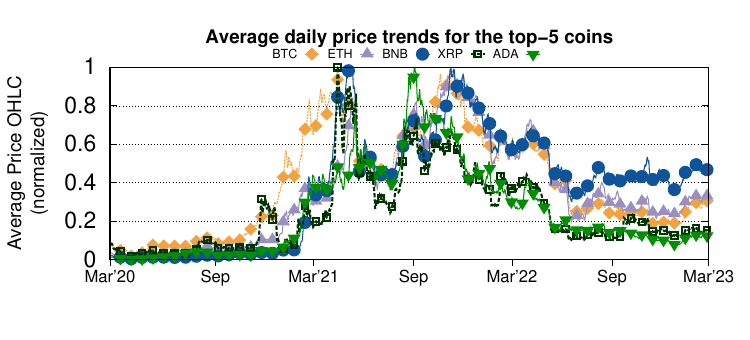}
	\end{center}
	\caption{Top-5 most-traded cryptocoins from March 1st, 2020 until March, 3rd, 2023. Dataset from Binance~\cite{binance}.\label{fig:avg-10}}
\end{figure}

Ownership of cryptocoins is stored in distributed ledgers (blockchains), along their transactions that record the transfers of coins across wallets.
Cryptocoins are exchanged, \ie, purchased and sold, for given trading prices.
History has shown the extreme volatility of such trading prices across all different sets of crypto-assets.
Reasons for such volatility include lack of adequate regulation, the inherent speculative nature of cryptocurrencies and the lack of a governmental or institutional guarantor, as well as \emph{pump-and-dump} actions enacted by large stakes (\ie, \emph{whales} owning large percentages of the issued coins).
Fig.~\ref{fig:avg-10} shows the normalized average OHLC (open-high-low-close) daily prices on Binance for the top-5 most-traded assets (\ie, BTC, ETH, BNB, XRP, ADA) since March 2020. 
Within short time-frames (\ie 6 months between March and September 2021), the prices vary sensibly.
Visually, one observes price variations following correlated trends, \eg an upward trend until March 2021, a short-term downward trend until June 2021, followed by an upward until September 2021. 
Co-movements persist also between March and June 2022, a time segment characterized by an overall price drop. 
It remains unclear the extent of the correlations between the considered cryptocoins and on which ones such correlations exist.
Coincidentally, major cryptocoin exchanges started to provide \emph{trend correlation} indicators to coin/wallet owners.
For example, Coinbase, defines the price correlation as \emph{the tendency of other asset prices to change at the same time as the asset shown on the page.} 
In their case, the correlation index represents the Pearson correlation $r$, leveraging the USD-based order books over the last 90 days (\url{https://bit.ly/3CsH076}). 
Given the enormous popularity of such digital assets among common public, facilitated by the easy access via mobile apps to these markets, such correlation indicators could drive end-users towards sell or buy decisions. 
It is thus important to fully understand such indicators and the degree of trust that end-users should dedicate to those. 

The nature of such correlations, as well as the evolution of the correlations through time, remain unexplored.
In addition, only a small subset of trend correlation indicators are typically offered, often at the discretion of the exchange.  
The first goal of this paper is to study the nature of these correlations, and unveil, via our study, their characteristics.
We leverage this correlation study to answer the following research questions.

\textit{\textbf{RQ1}: Exploiting cryptocoins correlations, can we forecast the trends of main coins (BTC/ETH) leveraging the trends of the altcoins?}

\textit{\textbf{RQ2}: What are the time series forecasting approaches most adapted to the cryptocoin market?}

Our \textbf{contributions} are as follows.
First, by leveraging the trading prices and other exchange metadata (\eg, open and closing price, market capitalization, volume) for a large set of 62 cryptocurrencies, we experimentally study the trend correlations between and across cryptocoins. 
We analyze daily, weekly and monthly correlation patterns of the two \emph{main coins}, BTC and ETH, against the remaining set of 60 cryptocurrencies (\emph{altcoins}) in our dataset.
Our study confirms strong correlations between the observed trends.
We select 14 highly-correlated altcoins (\ie, ADA, BAT, BNB, DASH, DOGE, LINK, LTC, NEO, QTUM, TRX, XLM, XMR, XRP, ZEC) to investigate causality relationships with BTC/ETH and study how state-of-the-art machine learning techniques (\eg GBMs, LSTM and GRU, detailed in \S\ref{sec:background}) can be used to forecast the main coins price series.

Our results show that all proposed models are able to provide reliable price forecasts (\ie, low RMSE - Root Mean Square Error) and to significantly beat constant mean/median regressors baselines.
We follow an \emph{open science} approach, releasing code and datasets at \url{https://github.com/quapsale/cryptoanalytics}. 

\textbf{Roadmap.} 
\S\ref{sec:background} covers background on cryptocoins and notions of time series analysis with machine-learning techniques.
\S\ref{sec:dataset} describes our datasets.
\S\ref{sec:correl} shows our correlation analysis. 
\S\ref{sec:causality} analyzes causality relationships among cryptocoins.
We present our experimental results in \S\ref{sec:eval}, related work in  \S\ref{sec:relwork} before concluding with our future work in \S\ref{sec:conclusion}.

%% file: background.tex
\section{Background}\label{sec:background}
\subsection{Cryptocoins in a Nutshell}\label{subsec:crypto}

Cryptocoins are digitally-encrypted assets, typically designed as full replacements for fiat currencies and used mostly in p2p networks.
Depending on the incentive nature of the underlying blockchain, cryptocoins (or token) are rewarded to nodes in the network.
We can differentiate between three main types of cryptocoins: \emph{(i)} Bitcoin and Ether, \emph{(ii)} altcoins, and \emph{(iii)} stablecoins.\footnote{Some characterizations define stablecoins as sub-classes of altcoins, together with secure tokens, utility tokens, and more. We leave as future work to study in-depth the correlations between such sub-types of altcoins.}
Altcoins are \emph{alternative} coins to maincoins, which we consider to represent Bitcoin and Ether in this paper.
Notable altcoins are Cardano (ADA), Litecoin (LTC) or Ripple (XRP). 
A stablecoin (\eg, Tether USDT, or USDC, \etc) is a class of cryptocurrency that attempts to offer price stability, backed by a reserve asset, \eg gold or the value of the American dollar. 

\subsection{Machine Learning and Time Series Analysis}\label{subsec:tsa}
A time series~\cite{shumway2000time,ts2008} is an $n$-tuple of observations collected sequentially over time, such as trends of interest rates and stock prices, daily high/low temperatures, the electrical activity of the heart, \etc.
A fundamental feature of time series is that adjacent observations are dependent~\cite{tsa1, tsa2}. 
The analysis of this dependence requires the development of a \textit{stochastic process}, a sequence of random variables $\{Y_t: t = 0, \pm 1, \pm 2, \pm 3,…\pm n\}$ used as a model for the time series.
An important class of stochastic models for describing time series are \textit{stationary processes}, where the probabilistic properties that govern the behavior of the process (\ie mean, variance and autocorrelation structure) are constant over time.
Typical stationary processes include \textit{white noise} processes, \ie a sequence of independent, identically distributed random variables.
In finance and economic systems, time series are typically \textit{nonstationary}, \ie without a constant mean over time. 
Such time series can exhibit deterministic trends, \ie cyclical fluctuations and seasonalities with periodical rises and falls. 
This behavior is observed also in cryptocoin time series, characterized by nonstationary trends and high volatility.

The purpose of time series analysis is generally twofold: \emph{(i)}~understand the mechanisms and the inner dynamics of an observed series, and \emph{(ii)}~forecast the future values of the series based on the historical ones.
The rapid development of ML techniques allowed new methodologies for time series modeling and forecasting, as a alternative to traditional econometrics approaches (\eg, ARMA and VAR processes~\cite{hamilton2020time}).
In this paper, we apply two families of such alternative models: \emph{gradient-boosting machines} (GBMs) and \emph{recurrent neural networks} (RNNs), both adapted to time series forecasting and described next in further details.

\textbf{Regression Trees.} Decision trees~\cite{ISLR} are a family of non-parametric ML algorithms, relevant for both regression and classification problems.
We focus on regression trees~\cite{loh2011classification} as they can predict the value of a target variable by learning simple decision rules inferred from the data.

Initially all the training observations are grouped in the same set, and subsequently split in $\{R_1, R_2, R_3, ..., R_n\}$ partitions according to a \emph{recursive binary splitting} (stop when each terminal node in the tree reaches a user-specified minimum size). 
For each new observation $x_j \in R_p$, the corresponding predicted value $\hat{y}_j$ will be the average of all the $m$ target variables $\{y_1, y_2,..., y_m\} \in R_p$. 
Compared to more complex ML algorithms, regression trees are easy to interpret and visualize, but their predictive power is weaker.

\textbf{Gradient-Boosting Machines.} One can overcome the limitations of regression trees by aggregating many decision trees with \emph{bagging} and \emph{boosting}, significantly improving their their predictive performance. 
The main idea behind these approaches is to improve a single weak model by combining it with other weak models in order to generate an \emph{ensemble}, \ie, a collective strong model. 

Bagging, or \emph{bootstrap aggregation}, is a statistical procedure that builds $n$ regression trees using $n$ bootstrapped training sets, aggregating (\ie, averaging) the resulting predictions. 
A specific case of bagging is represented by \emph{random forests}, where a sub-sample of $m$ predictors is chosen each time a tree split is considered from the bootstrapped training set. 

\begin{figure}[!t]
	\begin{center}
		\includegraphics[scale=0.45]{{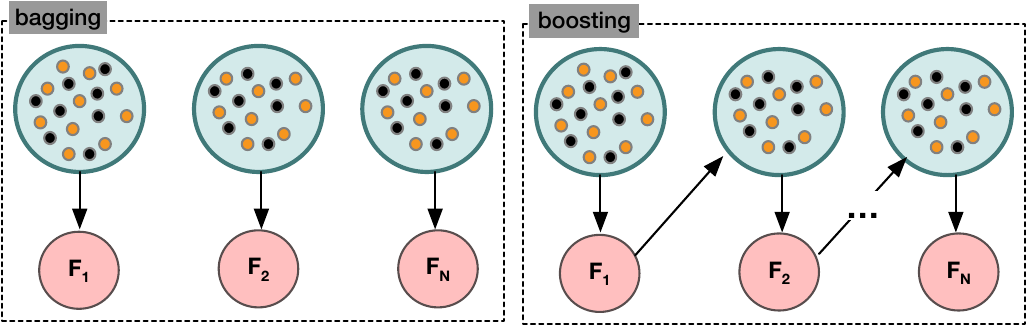}}
	\end{center}
	\caption{\label{fig:boost}Ensemble learning techniques: bagging and boosting.}
\end{figure}

Boosting does not involve bootstrap sampling: tree splits are formed sequentially on the original training set using information from previous trees (see Fig.~\ref{fig:boost}). 
Given a loss function, in each iteration a regression tree is ﬁt to the model and then added to the next iteration, in order to update the residuals (\ie the difference between the observed and the expected value). 
\emph{Gradient-boosting} extends traditional boosting where the iterative generation of weak models is determined minimizing the gradient over the chosen loss function.
For this reason, models built using this approach are known as \emph{gradient-boosting machines} (GBMs), or \emph{gradient-boosted trees} (GBTs). 
XGBoost, LightGBM and CatBoost are three notable state-of-the-art GBMs used later in this paper. 

XGBoost~\cite{xgboost} is an open-source, scalable and distributed GBM.
It builds trees in parallel rather than sequentially. 
It yields good results on a wide range of ML challenges (\eg, 17 out of 29 winning solutions on Kaggle's KDDCup 2015~\cite{xgboost}), and in non-ML applications, \eg energy physics and particle research~\cite{hbmc}. 
This is mainly due to its scalability across a diversity of scenarios, in addition to its capability to handle sparse data and efficiently approximate tree learning.

Microsoft's LightGBM~\cite{lightgbm} uses two novel techniques (\emph{gradient-based one-side sampling} and \emph{exclusive feature bundling}) to respectively deal with a large number of data instances and features. 
It is characterized by fast training speed and efficiency, fairly low memory usage and scalability.

CatBoost~\cite{catboost} is a GBM that introduces \emph{ordered boosting}, a modification
of the standard gradient boosting algorithm that avoids the \emph{prediction shift} of the learned model, a common problem that arises when training traditional GBMs.

We evaluate these three algorithms (XGBoost, LightGBM and CatBoost) with respect to their efficiency in forecasting cryptocoin time series, and compare them against neural network approaches, described next.
We show that gradient-boost approaches adjust more efficiently to the high variance observed in the cryptocoin time series.

\textbf{Deep Learning.} \emph{Neural networks} (NNs)~\cite{Goodfellow-et-al-2016} are ML models to approximate some function $f(x; \theta)$ by learning the value of the parameters $\theta$ that give the best approximation for $f$. 

These networks are typically represented by a chain of composed functions (\ie, \emph{layers} of the network), the first one being the input, the final one being the output and the middle ones called \emph{hidden layers}. 

Each layer in the network is characterized by an \emph{activation function} that describes how the respective output is propagated further to the next steps in the model, until it reaches the final output layer.
\begin{figure}[!t]
	\begin{center}
		\includegraphics[scale=0.5]{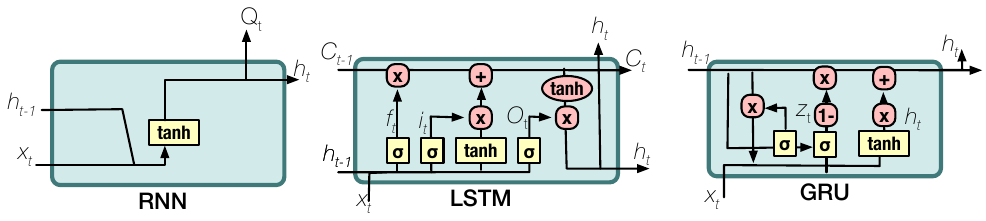}
	\end{center}
	\caption{\label{fig:rnns}Deep Learning approaches to time series forecasting: ``vanilla'' RNN, LSTM and GRU.}
\end{figure}

The overall number $n$ of layers gives the depth of the model.
The number $m$ of neurons for each layer represents the width. 
In real-world applications, $n$ and $m$ can be large~\cite{athlur2022varuna}.
For example, if the task is to model raw sensory data (\ie images) with a NN, the input layer represents the \emph{most visible} features of the object (\ie edges), while a large series of hidden layers will extract increasingly abstract features (\ie contours, corners) from the image~\cite{cnn}.

The training of a NN takes place by fitting the model hyperparameters using the gradient descent algorithm~\cite{rumelhart1985learning} over a user-specified loss function. 
The \emph{Stochastic gradient descent} (SGD algorithm) is a widely adopted technique where the overall gradient is estimated by taking the average gradient on mini-batches of training examples. 
A crucial parameter for the SGD algorithm is the learning rate.
Several techniques extend SGD~\cite{adagrad}.
\emph{Adaptive moment estimation} (Adam)~\cite{adam} is a variation of SGD used as optimizer to train our DL models (see \S\ref{sec:eval}).
It computes individual adaptive learning rates for different parameters from estimates of first and second moments of the gradients. 

\emph{Feed-forward neural networks} (FFNNs) are a family of NNs where the information flows through the function being evaluated from $x$, via the intermediate computations used to define $f$, and finally to the output $y$. 
In presence of feed-back connections in which outputs are fed back into themselves, NNs are called \emph{feed-back neural networks} (FBNNs) or, more commonly, \emph{recurrent neural networks} (RNNs), described next.

\textbf{Recurrent Neural Networks.} In RNNs, the behavior of hidden neurons is not only determined by the activations in previous hidden layers, but also by the activations at earlier times. 
The activation function for every hidden layer of a RNN is: $h^{(t)} = f(h^{(t-1)}, x^{(t)}, \theta)$. 
There, the hidden layer at the time $t$, $h^{(t)}$ is a function of the previous status, $h^{(t-1)}$, of the current input $x^{(t)}$ and of the activation function $\theta$.

The training process of RNNs is usually complex, due to the \emph{unstable gradient problem}: the gradient of the adopted cost function tends to get smaller or bigger as it is propagated back through layers, resulting in a final vanishing or exploding effect, respectively. 
As a result of this phenomenon, \emph{vanilla} RNNs are unable to model \emph{long term dependencies}, lacking predictive ability when dealing with long sequences of data.

\begin{figure}[!t]
	\begin{center}
		\includegraphics[scale=0.7,trim={0 16 0 10}]{{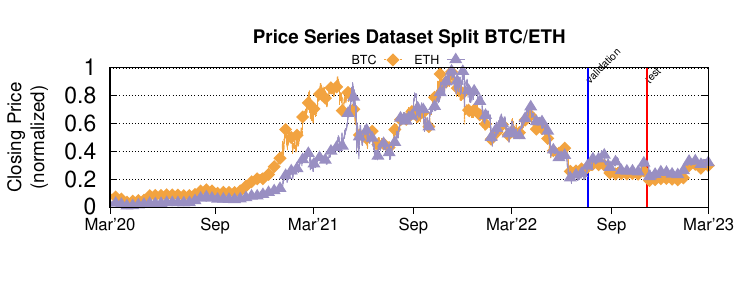}}
	\end{center}
	\caption{\label{fig:btc-vs-eth}Closing prices of BTC/ETH with validation/test split.}
\end{figure}	
\emph{Gated} RNNs circumvent this problem in practical applications, by means of \emph{long short-term memory} (LSTM) and \emph{gated recurrent unit} (GRUs). 
Fig.~\ref{fig:rnns} depicts these approaches.
LSTM~\cite{lstm} embed \emph{cells} with internal recurrence (a self-loop), in addition to the outer recurrence of the RNN. 
A LSTM cell is composed by \emph{(i)}~an input gate, \emph{(ii)}~a forget gate that controls the information flow and \emph{(iii)}~an output gate.
GRU~\cite{gru} uses a single gating unit that simultaneously controls the forgetting factor and the decision to update the state unit.

Deep learning models are also used in the context of time series forecasting~\cite{lim2021time,torres2021deep,shen2020novel} and financial scenarios~\cite{dingli2017financial,sezer2020financial}.
We contribute a comparative study of different deep-learning approaches specifically applied to cryptocoin  time series. 
Interestingly, a recent study~\cite{palamai} demonstrate how these price series are characterized by \textit{time-varying} and \textit{clustering} volatility (\ie large fluctuations in prices are followed by relatively large ones, and vice-versa), providing empirical evidence against the \textit{random walk} hypothesis in cryptocoins~\cite{fama}.

\subsection{Event-based Systems and Cryptocoin Price Series Forecasting}
The role of distributed and event-based systems in cryptocoin price forecasting is of key relevance. 
These systems provide the necessary infrastructure to collect and analyze real-time data streams, required to support accurate price predictions for the cryptocurrency markets as we show later.

The fluctuations in cryptocurrency markets produce a continuous stream of events (\ie price updates, trade volumes) from multiple sources, such as centralized as well as fully-decentralized (DEX) cryptocurrency exchanges. 
By distributing the collected data across multiple nodes, one can optimize the storage requirements and improve the accuracy of cryptocoin price predictions at scale.

Leveraging complex event processing techniques, one can unveil hidden patterns embedded in the price data, such as cross-correlation (\S\ref{sec:correl}) and causality (\S\ref{sec:causality}). 
In the context of cryptocoin price forecasting, it is possible to identify specific events, \ie sudden price movements, that can be exploited to analyze the overall market trend for a given cryptocurrency or among a broader set of coin assets. 

Distributed and event-based systems allow to process these event flows in real-time, allowing the ML forecasting models to adjust their predictions accordingly with a significant improvement in their performance. 
To efficiently model cryptocoin market trends, the ML models need to adapt quickly to the various and sudden fluctuations in market conditions by continously incorporating new information in real-time. 
By leveraging these infrastructures, is it possible to train state-of-the-art predictive ML models capable of providing accurate cryptocoin price forecasts even in extremely dynamic contexts like cryptocurrency markets.

%% file: dataset.tex
\section{Dataset Characterization}\label{sec:dataset}
We have two main objectives: \emph{(i)}~understand correlation patterns among cryptocoins and \emph{(ii)}~analyze causality relationships and forecast the price trends of the two main coins (BTC and ETH) basing on highly-correlated ($r \geq 0.6$) altcoins. 
The first analysis requires the use of a large dataset, including a significant number of cryptocurrencies.
For the second goal, we require a \emph{long} dataset, characterized by high granularity of time series variables. 
To achieve these goals, we leverage two distinct datasets, with records spanning 33 months, \eg from 20-02-2020 to 05-11-2022. Moreover, the second dataset contains additional records from 06-11-2022 to 26-02-2023, dedicated exclusively to the forecasting phase to avoid data leakage.

\textbf{CoinMarketCap dataset.} We collected the first dataset from CoinMarketCap~\cite{coinmarketcap}, a leading aggregator of cryptocurrency market data. 
From more than 9000 listed coins, we picked the top-62 for trading volumes. 
We exclude all stablecoins (\eg USDT, TUSD, DAI), as they are pegged to a reserve asset and do not follow the trend of other cryptocoins.
This dataset includes daily records for variables commonly used to study the time trends of financial instruments. 
Specifically, ``High'' and ``Low'' are the highest and lowest prices reached by the asset during the considered time frame; ``Open'' and ``Close'' the opening and closing market prices; finally, ``Volume'' indicates the quantity traded in the last period.
This dataset includes 61,380 observations, and a time series for each coin of 990 steps.
The two major coins in terms of traded volume are Bitcoin (BTC) and Ether (ETH), which we set as benchmarks for the remainder of study. 
As a matter of fact, the closing price trends of BTC and ETH over the considerd time window demonstrate on average an high positive correlation (Fig.~\ref{fig:btc-vs-eth}), and a correlation coefficient of the closing price $r \approx 0.9$ over the same time period.

\begin{figure*}[!t]
\centering
		\begin{subfigure}[b]{1.0\columnwidth}
		\includegraphics[scale=1.0, trim={ 110 0 0 0}]{{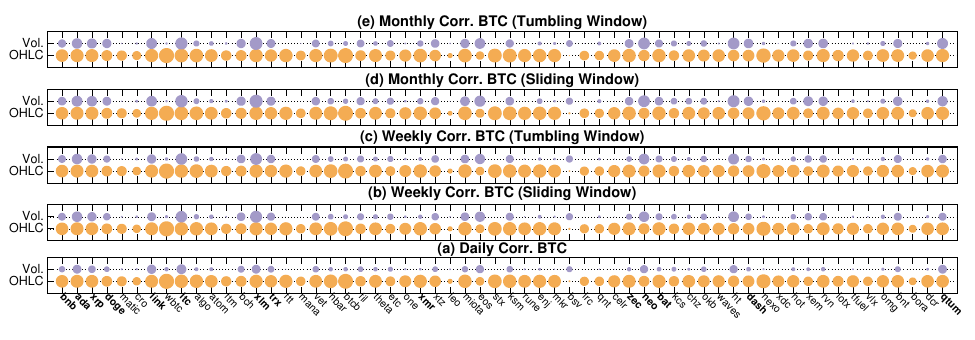}}
		\end{subfigure}

	\begin{subfigure}[b]{1.0\columnwidth}
		\includegraphics[scale=1.0,trim={ 110 0 0 0}]{{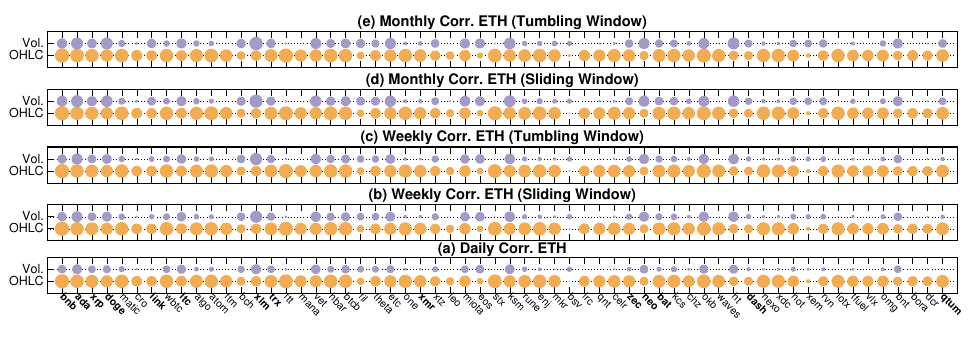}}		
	\end{subfigure}
	\caption{Correlogram between the average OHLC price and volume of 60 alt-coins and BTC (top) and ETH (bottom). Data from CoinMarketCap. \textbf{Bold}: coins used in \S\ref{sec:causality}.\label{fig:corr:ETH}}
\end{figure*}

\textbf{Binance dataset.} The second dataset, from Binance~\cite{binance}, presents higher-resolutoin data points (\ie, once per minute).
It includes the closing price records for the top-16 traded cryptocoins (\eg ADA, BAT, BNB, BTC, DASH, DOGE, ETH, LINK, LTC, NEO, QTUM, TRX, XLM, XMR, XRP, ZEC). 
The dataset consists of 25,218,800 observations, and each coin's time series includes 1,576,175 steps.

%% file: correl.tex
\section{Correlation Analysis}\label{sec:correl}
We use the CoinMarketCap dataset to identify cross-correlation patterns in cryptocurrency trends. 
We represent those correlations, averaged among all the studied variables (\ie, High, Low, Open, Close as an average OHLC, and volume), as a series of \emph{cross-correlograms} of coins, for BTC (Fig.~\ref{fig:corr:ETH}-top) and ETH (Fig.~\ref{fig:corr:ETH}-bottom).

We analyze the correlations of 60 altcoins present in our dataset against BTC and ETH.
To understand the extent in time of the correlations, we consider three different time frames: daily, weekly and monthly correlations.
For weekly and monthly correlations, we define the sequence segments adopting a sliding window approach. Observations are grouped within a window that slides across the data stream.
These are shown for BTC and ETH, respectively in Fig.~\ref{fig:corr:ETH}-top(b/d) and Fig.~\ref{fig:corr:ETH}-bottom(b/d).
Further, we include the study of correlations over \emph{tumbling} windows, where there is no overlapping of data clusters, for BTC in Fig.~\ref{fig:corr:ETH}-top(c/e) and for ETH in Fig.~\ref{fig:corr:ETH}-bottom(c/e).
 
The daily observations for each coin are averaged over sliding partitions of 7 and 30 days, respectively. 
The correlations with other coins are computed on the resulting aggregated values. 
The radius of each circle represents the strength of the correlation (\ie, the Pearson coefficient $r$) between the current altcoin and the corresponding main coin (\eg BTC or ETH).

The cross-correlograms show that vast majority of the considered altcoins are strongly correlated with the two market leaders, \ie their average values of Pearson coefficient is close to $1$, with BSV (an altcoin issued from a recent hard fork of the Bitcoin Cash blockchain) being one notable exception (\ie, the weakest correlation). 

The tumbling and sliding correlations follow very similar patterns: the correlation strengths increase if compared to the daily patterns (Fig.~\ref{fig:corr:ETH}), with strong similarity between the ETH and BTC correlograms.

\begin{figure*}[!t]
	\begin{center}
		\includegraphics[scale=0.35]{{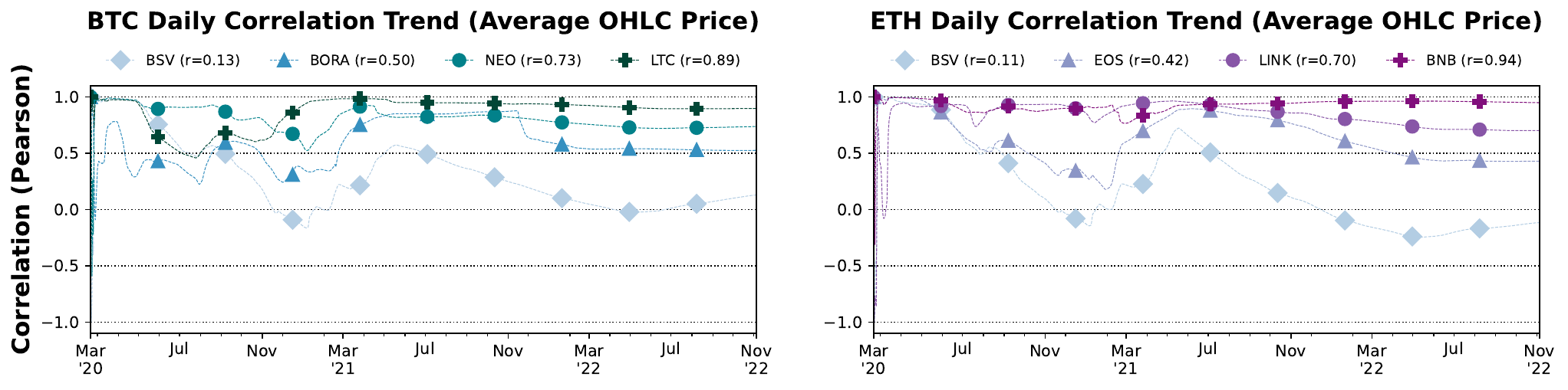}}
	\end{center}
	\caption{\label{fig:btc-corr-OHLC}Daily cross-correlation trends for 4 differently correlated alt-coins against BTC (left) and ETH (right).}
\end{figure*}

Fig.~\ref{fig:btc-corr-OHLC} provide a deeper insight on the daily evolution of average OHLC price correlations (as the Pearson coefficient $r$) between BTC or ETH and a set of four alt-coins. 
Each alt-coin represents a different class of correlation intensity: \textit{weak} (0 to 0.25), \textit{average} (0.25 to 0.5), \textit{above average} (0.5 to 0.75) and \textit{high} (0.75 to 1). 
To represent the correlation trend over time, we compute the Pearson coefficient $r(A,B)_{t_n}$ between a main coin A and an alt-coin B at each time step $t_n$ considering the distribution of prices for A and B in the previous $n-1$ steps. 
At the beginning of the time series, we set conventionally $r(A,B)_{t_0} = 1$.

The correlation trend of BTC with NEO and LTC is stable and always very close to 1.
Exceptions include BSV and BORA, with a standalone trend. 
It appears to be a seasonality trend in the Pearson coefficient distribution between BTC and BSV/BORA, with a common decrease in Fall 2020 and Winter 2021 (\ie, August to February), followed  by an increase in Spring 2021 (\ie March 2021 onward). 
From there on, the correlation trend for BORA starts to align with NEO and LTC, while the BSV trend starts to rapidly decrease from June 2021. 

For ETH, the correlation with EOS, LINK and BNB is stable and very close to 1.
Two exceptions exist: the EOS trend between Fall 2020 and Spring 2021 and a general divergence from December 2021 onward. 
Similar to BTC, the correlation between ETH and BSV shows a possible seasonal trend with a decrease between August 2020 and February 2021, followed by a stark increase during Spring (\ie March to May 2021) and a subsequent decrease from June 2021 onward.

We select 14 altcoins (Fig.~\ref{fig:corr:ETH}) for the analysis of causality patterns, to identify the predicting variables for our models. 
Such 14 altcoins include, in our Binance dataset, finer timestamp resolutions and strong correlations ($r \geq 0.6$).
We leave as future work to exploit even higher (\ie milliseconds) resolution datasets.

%% file: causality.tex
\section{Causality Analysis}\label{sec:causality}
The presence of strong correlation patterns between altcoins and main coins is not sufficient to deduce a cause-and-effect relationship. 
Therefore, we investigate the Granger causality~\cite{granger} between a set of 14 correlated altcoins (ADA, BAT, BNB, DASH, DOGE, LINK, LTC, NEO, QTUM, TRX, XLM, XMR, XRP, ZEC) and BTC/ETH.
To do so, we use a modified version of the Binance dataset where the original per-minute observations are averaged over a daily window. 

\begin{table}[!t]
	\centering
	\setlength{\tabcolsep}{12pt}
\begin{center}
		\rowcolors{1}{gray!0}{gray!10}

\begin{tabularx}{\columnwidth}{lrrr}
			\rowcolor{gray!25}
			$\mathbf{H_0}$ & $\mathbf{\chi^2}$ & \textbf{p-value} & \textbf{Result} \\
			\rowcolor{gray!1}
			\texttt{ADA} $\centernot \implies$ \texttt{BTC}   &   31.31 & 0.018 &  \xmark \\
			\rowcolor{gray!10}
			\texttt{BAT} $\centernot \implies$ \texttt{BTC}    & 83.98 & 5.8e-5 &  \xmark\\
			\rowcolor{gray!1}
			\texttt{BNB} $\centernot \implies$ \texttt{BTC}    & 5.29 & 0.15 &  \cmark\\
			\rowcolor{gray!10}
			\texttt{DASH} $\centernot \implies$ \texttt{BTC}   &  19.06 & 0.014 &  \xmark \\
			\rowcolor{gray!1}
			\texttt{DOGE} $\centernot \implies$ \texttt{BTC} & 69.43 &  0.001  &  \xmark\\
			\rowcolor{gray!10}
			\texttt{LINK} $\centernot \implies$ \texttt{BTC}  & 8.85 & 0.012 & \xmark  \\
			\rowcolor{gray!1}
			\texttt{LTC} $\centernot \implies$ \texttt{BTC}  & 120.43  & 9.2e-11 & \xmark  \\
			\rowcolor{gray!10}
			\texttt{NEO}  $\centernot \implies$ \texttt{BTC}  & 55.82  & 5.4e-5 & \xmark  \\
			\rowcolor{gray!1}
			\texttt{QTUM} $\centernot \implies$ \texttt{BTC}   & 59.48  & 2.6e-5 & \xmark  \\
			\rowcolor{gray!10}
			\texttt{TRX} $\centernot \implies$ \texttt{BTC}  & 84.77  & 5.2e-6 & \xmark  \\
			\rowcolor{gray!1}
			\texttt{XLM} $\centernot \implies$ \texttt{BTC}  & 69.43  & 0.001 & \xmark  \\
			\rowcolor{gray!10}
			\texttt{XMR} $\centernot \implies$ \texttt{BTC}  & 46.69  & 0.027 & \xmark  \\
			\rowcolor{gray!1}
			\texttt{XRP} $\centernot \implies$ \texttt{BTC}  & 77.05  & 3.9e-4 & \xmark  \\
			\rowcolor{gray!10}
			\texttt{ZEC} $\centernot \implies$ \texttt{BTC}  & 47.14  & 3.7e-7 & \xmark  \\
			
			\rowcolor{gray!50}
			\rowcolor{gray!25}
\bottomrule
\end{tabularx}		
\begin{tabularx}{\columnwidth}{lrrr}
		\rowcolor{gray!25}
		$\mathbf{H_0}$ & $\mathbf{\chi^2}$ & \textbf{p-value} & \textbf{Result} \\
		\rowcolor{gray!1}
		\texttt{ADA} $\centernot \implies$ \texttt{ETH}  & 89.75  & 2.2e-8 & \xmark  \\
		\rowcolor{gray!10}
		\texttt{BAT} $\centernot \implies$ \texttt{ETH}   & 91.27 & 4.2e-8 & \xmark \\
		\rowcolor{gray!1}
		\texttt{BNB} $\centernot \implies$ \texttt{ETH}   & 18.76 & 0.04 & \xmark \\
		\rowcolor{gray!10}
		\texttt{DASH} $\centernot \implies$ \texttt{ETH}  &  70.56 & 1.4e-7 & \xmark \\
		\rowcolor{gray!1}		
		\texttt{DOGE} $\centernot \implies$ \texttt{ETH} & 134.51 &  2.7e-13  &  \xmark\\
		\rowcolor{gray!10}		
		\texttt{LINK} $\centernot \implies$ \texttt{ETH}  & 23.53 & 0.002 & \xmark  \\
		\rowcolor{gray!1}
		\texttt{LTC} $\centernot \implies$ \texttt{ETH}  & 65.20  & 1.1e-6 & \xmark  \\
		\rowcolor{gray!10}
		\texttt{NEO}  $\centernot \implies$ \texttt{ETH}  & 142.76  & 1.8e-13 & \xmark  \\
		\rowcolor{gray!1}
		\texttt{QTUM} $\centernot \implies$ \texttt{ETH}   & 117.14  & 5.7e-15 & \xmark  \\
		\rowcolor{gray!10}
		\texttt{TRX} $\centernot \implies$ \texttt{ETH}  & 41.60  & 0.001 & \xmark  \\
		\rowcolor{gray!1}
		\texttt{XLM} $\centernot \implies$ \texttt{ETH}  & 134.51  & 2.7e-13 & \xmark  \\
		\rowcolor{gray!10}
		\texttt{XMR} $\centernot \implies$ \texttt{ETH}  & 73.65  & 1.6e-5 &  \xmark  \\
		\rowcolor{gray!1}
		\texttt{XRP} $\centernot \implies$ \texttt{ETH}  & 145.04  & 7.9e-14 & \xmark  \\
		\rowcolor{gray!10}
		\texttt{ZEC} $\centernot \implies$ \texttt{ETH}  & 140.84  & 1.2e-14 & \xmark  \\
		
		\rowcolor{gray!50}
		\rowcolor{gray!25}
\bottomrule
 \end{tabularx}
 \end{center}	
	\caption{\label{tab:granger-btc}Results of the m-Wald test for Granger causality in Bitcoin (top) and Ether (bottom) price series. Legend: $\mathbf{H_0}$ rejected (\xmark) or not rejected (\cmark). $\mathbf{H_0}$ rejected if $\mathbf{p \leq 0.05}$.}
\end{table}

We leverage Toda-Yamamoto (T-Y)~\cite{todayamamoto} to estimate the one-directional Granger causality between a coin pair:
$A_c \implies M_c$,
where $A_c \in \{\text{ADA, BAT, BNB, ..., ZEC}\}$ is an altcoin and $M_c \in \{\text{BTC, ETH}\}$ is a main coin. 
We focus on analyzing whether $A_c$ can be used as a predictor for $M_c$, since the existence of an opposite relationship is mostly evident.
According to T-Y, we test the null hypothesis that $A_c$ does not Granger-cause $M_c$:
$H_0: A_c \centernot \implies M_c$.
Given $d$ the maximum integration order among the two series, we generate an optimized VAR($p+d$) model from the two coins considered and test the hypotheses that the coefficients of the first $p$ lags of $A_c$ are zero in the $M_c$ equation (\ie $A_c$ does not Granger-cause $M_c$). 
We then apply a modified Wald (m-Wald) \cite{wald} test that follows a $\chi^2$ distribution with $p$ degrees of freedom.

To find $d$, we followed an iterative approach testing each coin series for stationarity with the \sloppy{Kwiatkowski–Phillips–Schmidt–Shin~\cite{kwiatkowski1992testing}} (KPSS) test and the Augmented Dickey–Fuller (ADF)~\cite{dickey1981likelihood} test. 
For all cryptocurrencies, we found out that $d=1$, meaning that only first-order differencing was needed to make the series stationary. 
A similar approach was followed also to find $p$, setting up different VAR models for each coin pair and choosing the one with the minimum Akaike Information Criterion (AIC) \cite{Akaike1998}.

Table~\ref{tab:granger-btc} resumes the m-Wald test results. 
With the exception of BNB, all considered altcoins Granger-cause BTC. 
Moreover, all the altcoins Granger-cause ETH. 

We exploit these causality relationships by using the altcoins that Granger-cause BTC/ETH as feature (\ie predictor) variables to forecast Bitcoin and Ether closing prices (see \S\ref{sec:eval}).

%% file: eval.tex
\section{Evaluation}\label{sec:eval}
This section presents our experimental evaluation. 
Specifically, we answer \emph{RQ1} and forecast BTC and ETH closing prices based on the behavior of a set of correlated altcoins price series. 
In \S\ref{sec:causality} we analyzed Granger-causality to identify the altcoins to use as predictor variables: these are, respectively, ADA, BAT, DASH, DOGE, LINK, LTC, NEO, QTUM, TRX, XLM, XMR, XRP, ZEC for Bitcoin and ADA, BAT, BNB, DASH, DOGE, LINK, LTC, NEO, QTUM, TRX, XLM, XMR, XRP, ZEC for Ether. 
Then, after examining the predictive ability of our models, we address \emph{RQ2} by describing the most adapted ML-based forecasting approaches to cryptocoins price trends.

\begin{table}[!t]
	\centering
	\scriptsize
	\rowcolors{1}{gray!10}{gray!0}
		\begin{tabularx}{\columnwidth}{Xrrrr}
		\rowcolor{gray!50}
		\textbf{} &  \multicolumn{2}{c}{\textbf{Learning Rate}} & \multicolumn{2}{c}{\textbf{Iterations}}\\
		\rowcolor{gray!25}
		\textbf{Model} & \textbf{BTC} & \textbf{ETH} & \textbf{BTC} & \textbf{ETH} \\
		\rowcolor{gray!10}
		XGBoost  &   0.01 & 0.01 &  52 & 384 \\
		LightGBM   &   0.01 & 0.1 &  39 & 52 \\
		CatBoost   &   0.1 & 0.1 &  5 & 24 \\
		LSTM   &   0.001 & 0.001 &  3 & 3 \\
		GRU   &   0.001 & 0.001 &  4 & 32\\
		\bottomrule
	 \end{tabularx}
	\caption{\label{tab:gbm}Optimal model hyperparameters in BTC and ETH price series modeling.}
\end{table}

\textbf{Experimental setup.} For model training, we use Ubuntu 22.04.2, Linux kernel 5.19.0-35-generic, 8-core AMD Ryzen\texttrademark{} 7 5700G clocked at 3.8 Ghz, 32\,GB RAM and and NVIDIA GeForce RTX\texttrademark{} 3070.
For RNNs, we rely on PyTorch 1.11.0 (with CUDA 11.3) and PyTorch Lightning 1.5.10.
Finally, the GBMs are implemented on top of XGBoost 1.5.1, LightGBM 3.3.2 and CatBoost 1.0.5.

\textbf{Training and validation.} We split the Binance dataset in three parts (see Fig.~\ref{fig:btc-vs-eth}): a training set containing 1,260,941 observations for each price series (\ie, from 2020-02-20 08:06:00 to 2022-07-19 02:05:00), a validation set with 157,617 observations (\ie, up to 2022-11-05 13:02:00) and a final test set with the remaining 157,617 observations (\ie, until to 2023-02-26 23:59:00), used only for the prediction phase. 

\begin{figure*}[!t]
	\begin{center}
		\includegraphics[scale=0.35]{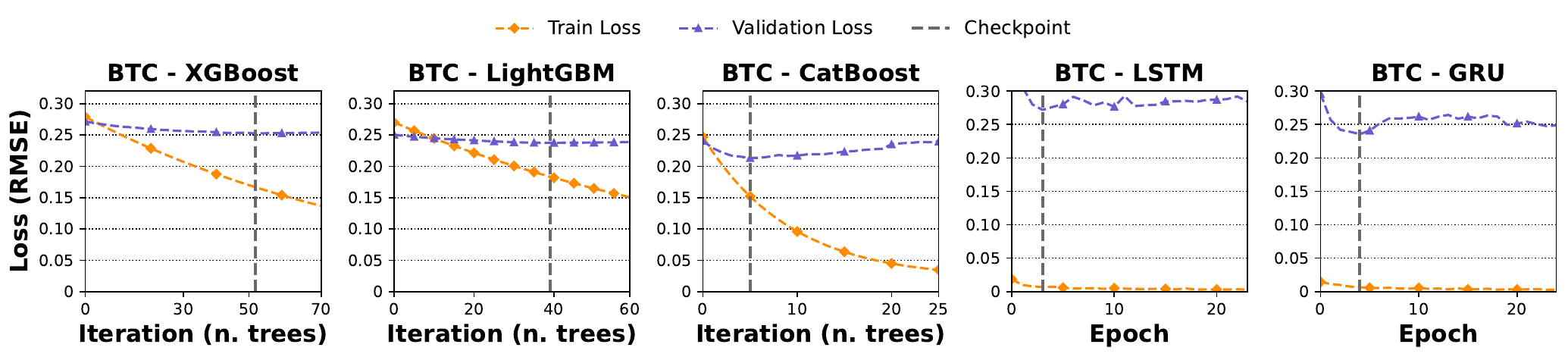}
		\includegraphics[scale=0.35]{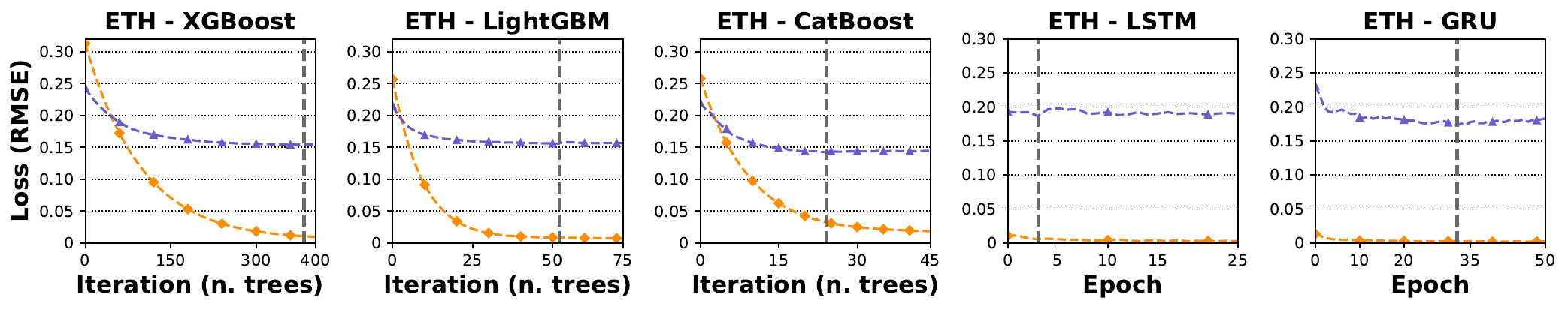}
	\end{center}
	\caption{\label{fig:eth-gbm-training}Model train/validation loss trend for BTC and ETH closing price forecast.}
\end{figure*}

We first examined the behavior of three gradient-boosting machines (\eg, XGBoost, LightGBM and CatBoost). 
Those models are fast to train, with only two hyperparameters: learning rate and the optimal number of trees/iterations. 
Table~\ref{tab:gbm} summarizes the optimized final versions, with the respective hyperparameters used to model Bitcoin and Ether price series.
We used the mean squared error (MSE) as the loss function for the model learning process.
To find the optimal learning rate $lr$, we performed a grid search over a list of 6 possible values (0.000001, 0.00001, 0.0001, 0.001, 0.01, 0.1) by cross-validating each model on 10 folds generated from the original data.
To fine-tune the number of trees, we set an initially high upper limit of 500 iterations and then early-stopped the training after 20 consecutive rounds with no improvements in the validation set loss. 
Fig.~\ref{fig:eth-gbm-training} shows the learning process for Bitcoin and Ether by highlighting the trend of train and validation losses respect to the number of trees used. 
After a sufficient number of iterations (average of 32 for BTC and of 153 for ETH), the validation loss became flat, reaching an optimal point (\ie the \emph{checkpoint}) that indicates the final model version with the lowest validation loss.
We use such models during the time series prediction phase.

Next, we studied the performance of two recurrent neural networks, \ie, LSTM and GRU. 
Compared to GBMs, the training process for those models is slower in consideration of the high number of internal parameters (weights and biases) to fit. 
We used the MSE as the loss function, and the adaptive moment estimation (Adam) algorithm as the learning process optimizer. We set for both models $n=10, m=100$, where $n$ is the number of hidden layers (network length) and $m$ the number of neurons per layer (model width).
For both LSTM and GRU we fine-tuned the initial learning rate $lr$ 

and the number of iterations (epochs) required for the training.
To find $lr$, we launched a small run starting from an initial value of 0.000001 and increasing it after each processed batch.

Table~\ref{tab:gbm} summarizes the optimized final versions of LSTM and GRU with the respective hyperparameters used to model Bitcoin and Ether price series.

\begin{figure*}[!t]
    \centering
	\begin{center}
		\includegraphics[scale=0.35]{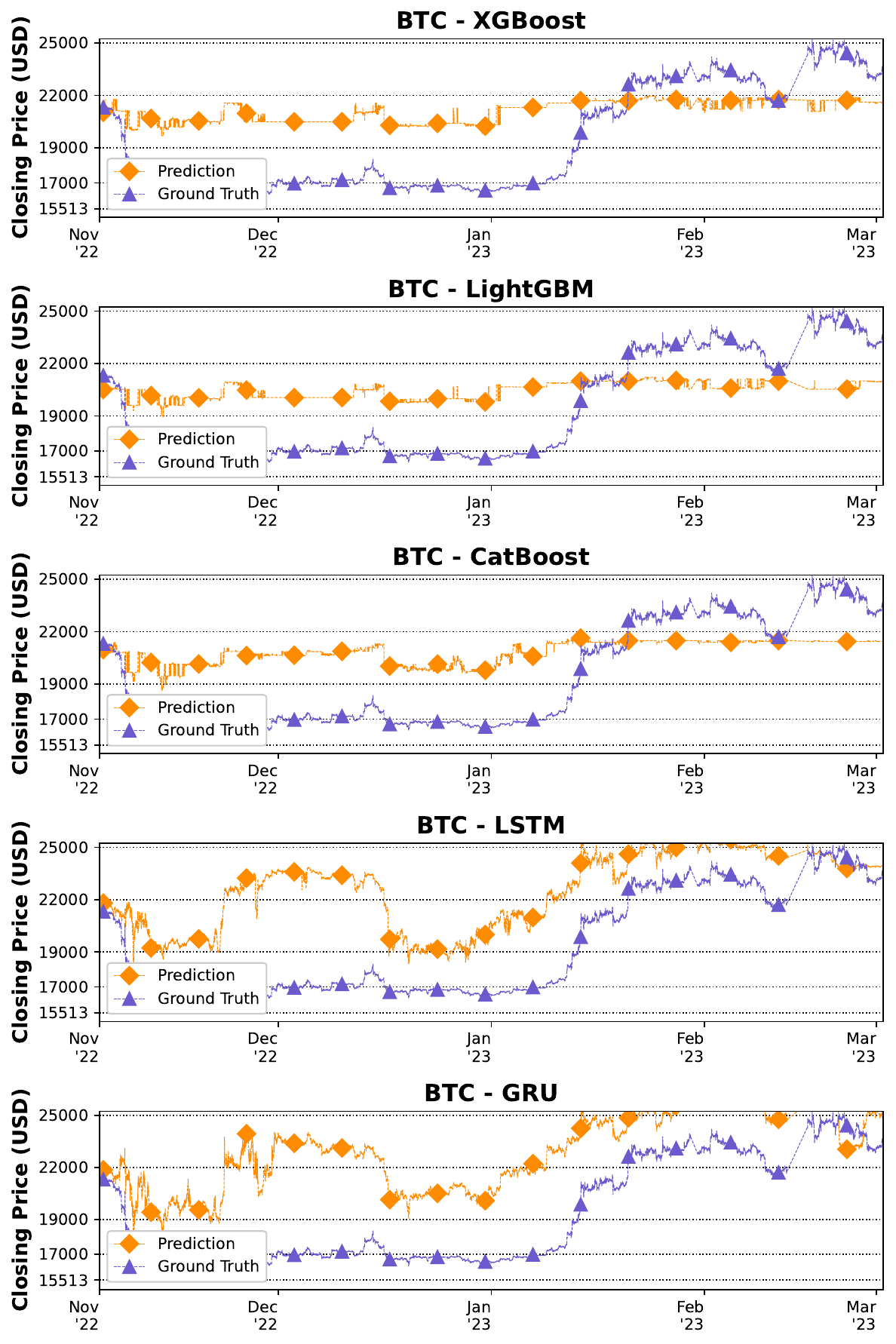}
		\includegraphics[scale=0.35]{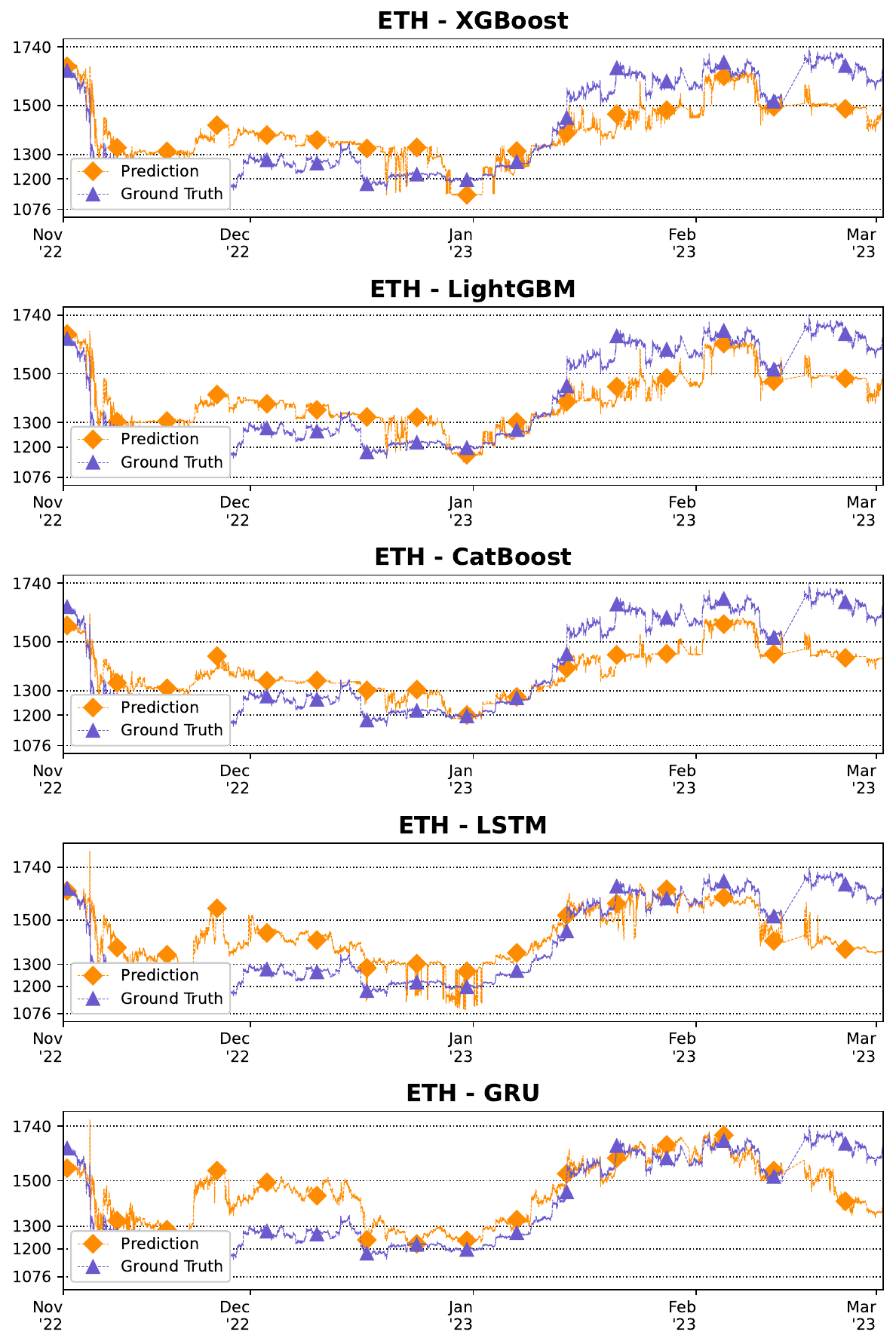}
	\end{center}
	\caption{\label{fig:prediction}Model forecasts for closing price. Left: BTC. Right: ETH.}
\end{figure*}

To fine-tune the number of training epochs, we set an initial limit of 500 iterations, and then early-stopped the training after 20 consecutive rounds with no improvements in the validation set loss. 
Given our machine specifications, we set a batch size of 256 training samples. 
Fig.~\ref{fig:eth-gbm-training} shows the training process for Bitcoin and Ether, by highlighting the trend of train and validation losses against the number of learning epochs.
The validation loss rapidly reached the optimum (\ie, \emph{checkpoint}), before flattening or overfitting as the number of training epochs increases. 
Here the patterns detected in the closing price trends are not complex enough to require an excessive number of training steps.
As for the GBMs, the checkpoint represents the optimal trained version of the the model, used later on for forecasting.

\textbf{Price series forecasting.}
Comparing the forecasts against the ground truth, all models achieve reliable estimations of the original price series for BTC and ETH. 

The gradient-boost machines (XGBoost, LightGBM and CatBoost) predict with high accuracy either \emph{stable} trends, for which there is no trace of short-term peaks/falls, and more unstable ones. 
From Fig.~\ref{fig:prediction}, forecasts from the three GBMs expose stable distributions with few, localized drifts, in particular nearby more increased shifts in the predicted price series. 
Instead, LSTM and GRU are less accurate in modeling stable trends, as shown by the price predictions for BTC between December 2022 and February 2023.

%% file: discussion.tex
\textbf{Comparison against statistical metrics.}\label{sec:discussion}
Finally, we evaluate how the given forecasting models compare against four statistical metrics: MSE, root mean square error (RMSE), mean absolute error (MAE) and mean absolute percentage error (MAPE). 
We include results against two deterministic baselines: \emph{(i)} a constant mean regressor (\ie always predicts the average closing price) and \emph{(ii)} a constant median regressor (\ie always predicts the median closing price).
Table~\ref{tab:disc-eth} reports these results for BTC and ETH.
All models significantly outperform the deterministic baselines, GBMs being the best (LightGBM for BTC and XGBoost for ETH). 
This is expected: as seen in \S\ref{sec:eval}, these models are efficient in modeling both long-term dependencies trends and short period drifts. 

For BTC, considering the RMSE, the average value between the three GBMs is 3050, \ie a 22\% improvement respect to the RNNs (average of 3936) and 74\% improvement against the two baselines.
For ETH, the average RMSE between the three GBMs is 120, that is a 14\% improvement respect to the RNNs (average of 140) and a 85\% improvement respect to the baselines.

\begin{table}[!t]
	\centering
	\scriptsize
    \setlength{\tabcolsep}{3pt}
	\begin{center}
		\rowcolors{1}{gray!10}{gray!0}
		\begin{tabularx}{\columnwidth}{lrrrrrrrr}
			\rowcolor{gray!50}
			\textbf{} &  \multicolumn{2}{c}{\textbf{MSE}} & \multicolumn{2}{c}{\textbf{RMSE}} & \multicolumn{2}{c}{\textbf{MAE}} & \multicolumn{2}{c}{\textbf{MAPE}} \\
			\rowcolor{gray!25}
			\textbf{Model} & \textbf{BTC} & \textbf{ETH} & \textbf{BTC} & \textbf{ETH} &  \textbf{BTC} & \textbf{ETH} & \textbf{BTC} & \textbf{ETH} \\
			\rowcolor{gray!10}
			\textbf{XGBoost}  &   10,035,762 &  \textbf{13,891} & 3,167 & \textbf{118} & 2,859 & \textbf{102} & 0.16 & \textbf{0.07} \\
			\textbf{LightGBM}   & \textbf{8,865,905} & 14,321 &  \textbf{2,977} & 120 & \textbf{2,768} & 102 & \textbf{0.15} & 0.07 \\
			CatBoost   &  9,036,512 &  15,127 & 3,006 & 123 & 2,746 & 104 &  0.15 &  0.07 \\
			LSTM   &  14,604,921 &  20,613 &  3,821 &  144 & 3,319 & 117 & 0.18 & 0.08 \\
			GRU   &  16,415,682  &  18,167 &  4,051 &  135 & 3,735 & 106 & 0.21 & 0.08 \\
			Mean Regr.   & 137,296,033 & 197,503 &  11,717 &  444 & 11,330 & 401 & 0.63 & 0.31 \\
			Median Regr.   & 138,962,435 & 122,789 & 11,788 & 350 & 11,403 & 295 & 0.63 & 0.24 \\
			\bottomrule
		\end{tabularx}
	\end{center}
	\caption{\label{tab:disc-eth}Comparison of MSE, RMSE, MAE and MAPE for BTC and ETH price forecasting models (in bold: best models).}
\end{table}

%% file: relwork.tex
\section{Related Work}
\label{sec:relwork}
The study of co-movement and cross-correlation events in cryptocurrency market trends has recently attracted the interest of academia.

Katsiampa~\cite{KATSIAMPA2019221} investigated the volatility dynamics of the two main coins, (\eg BTC and ETH), finding evidence of interdependencies between the two and price responsiveness to major news in the market.
In~\cite{ASLANIDIS2019130}, authors showed that cryptocurrencies exhibit similar mean correlation among them, with an unstable trend over time.
The same study computed the correlation between four cryptocoins (BTC, DASH, XMR and XRP) and three traditional assets (S\&P 500 Composite, S\&P US Treasury bond and Gold Bullion LBM), detecting an independent behavior respect to other financial markets.
\cite{KUMAR20190712} identify Bitcoin as cryptocurrency market leader using wavelet-based methods, showing how other coins trends are dependent from BTC price movements.
Hence, Bitcoin price drops are immediately reflected in other cryptocurrency prices.
\cite{CHAUDHARI2020101130} studied the global behavior for the cryptocurrency market discovering distinct and not time-persistent community structures characterized by cross-correlation.

Researchers directly compared statistical methods and deep-learning methods in the context of time series forecasting.
In~\cite{siami2018comparison}, LSTM is compared with autoregressive integrated moving average methods (ARIMA), confirming the superiority of LSTM, as we also have shown.

In literature, AutoML/AutoAI methods have been used for building time series prediction models in automated manner. For example, \cite{automl} introduced a zero configuration system to automatically train, optimize and choose the best forecasting approach among various classes of models for a given dataset.

Rather than exploiting correlation relations, \cite{kraaijeveld2020predictive} exploits social network activities (\ie Twitter) to predict the price trends of cryptocoins.
Authors use 9 cryptocoins and demonstrate the predictive power for a small subset (BTC, BCH and LTC).

%% file: conclusion.tex
\section{Conclusion and Future Work}\label{sec:conclusion}
Cryptocoins expose very volatile trends.
Despite, correlations exist between altcoins and maincoins (\eg Bitcoin and Ether), suggesting it might be possible to forecast maincoins evolution by means of altcoins. 

We presented a correlation study between BTC, ETH and 60 other altcoins, showing the existence of strong correlations.

Then, we investigated Granger-causality relationships between altcoins and main coins in order to forecast their price trends by using machine learning approaches for time-series forecasting.
We leveraged several gradient-boosting machines (\ie XGBoost, LightGBM, CatBoost) and recurrent neural networks (\ie LSTM, GRU).
These techniques show good forecasting results, well suited for the cryptocoin markets.

Our work is a significant starting point for further analyses in co-movement behaviors within cryptocoin markets and in modeling and forecasting trends of the asset prices. 
We will extend our study to different application domains, such as e-health (\ie, to forecast possible health anomalies by observing a large set of body signals captured by body sensors) and environmental scenarios (\ie earthquake predictions).
We will look into correlations between cryptocoins and more classical financial markets. 
Our preliminary analysis in this direction indicates high correlation between BTC and the NASDAQ index ($r = 0.91$) and poor correlation with FTSE ($r = 0.19$).

Datasets and code for reproducibility are available at \url{https://github.com/quapsale/cryptoanalytics}.

\balance